\documentclass[11pt]{article}
\usepackage{amsfonts,amsmath,amssymb,amsthm}
\usepackage[pdftex]{graphicx}
\usepackage[hmargin=1in,vmargin=1in]{geometry}
\usepackage{natbib}
\usepackage[colorlinks,citecolor=blue]{hyperref}
\usepackage{enumerate}
\usepackage{caption}
\usepackage[usenames,dvipsnames]{xcolor}
\usepackage{footmisc,fancyvrb} 
\usepackage{xfrac} 
\usepackage{cleveref}
\usepackage{multirow}
\usepackage{threeparttable}
\usepackage{array}
\newcolumntype{C}[1]{>{\centering\let\newline\\\arraybackslash\hspace{0pt}}m{#1}}
\newcolumntype{L}[1]{>{\raggedright\let\newline\\\arraybackslash\hspace{0pt}}m{#1}}

\VerbatimFootnotes
\def\qed{\rule{2mm}{2mm}}

\usepackage{tikz,graphicx,pgfplots,pgfkeys,subcaption}   

\def\addlegendimage{\csname pgfplots@addlegendimage\endcsname}

\usepackage[pagewise,mathlines]{lineno}
\mathchardef\dash="2D

\theoremstyle{definition}

\newtheorem{corollary}{Corollary}[section]
\newtheorem{proposition}{Proposition}[section]

\newtheorem{assump}{Assumption}

\newtheorem{remark}{Remark}[section]
\usepackage{etoolbox} 
\AtEndEnvironment{remark}{~\qed}%
\AtEndEnvironment{example}{~\qed}%

\begin{document}

\author{
Vishal Kamat \\
Toulouse School of Economics\\
University of Toulouse Capitole\\
\url{vishal.kamat@tse-fr.eu} 
}

\title{On the Identifying Content of Instrument Monotonicity \footnote{I thank Yuehao Bai, Ivan Canay, Azeem Shaikh and Alex Torgovitsky for helpful discussions and comments. Funding from ANR under grant ANR-17-EURE-0010 (Investissements d'Avenir program) is gratefully acknowledged.}}

\maketitle

\begin{abstract}
This paper studies the identifying content of the instrument monotonicity assumption of \cite{imbens/angrist:94} on the distribution of potential outcomes in a model with a binary outcome, a binary treatment and an exogenous binary instrument. Specifically, I derive necessary and sufficient conditions on the distribution of the data under which the identified set for the distribution of potential outcomes when the instrument monotonicity assumption is imposed can be a strict subset of that when it is not imposed.
\end{abstract}



\newpage

\section{Introduction}\label{sec:introduction}

This paper studies the identifying content of the instrument monotonicity assumption of \cite{imbens/angrist:94} on the distribution of potential outcomes in a model with a binary outcome, a binary treatment and an exogenous binary instrument---see \cite{swanson/etal:18} for a recent overview of a number of identification results developed in the context of such a model. Specifically, let $Y$ denote the observed binary outcome and $D$ denote an indicator for whether treatment was received or not. Further, let $Y_0$ denote the binary potential outcome if treatment was not received and $Y_1$ denote the binary potential outcome if treatment was received. The observed and potential outcomes are related by the equation
\begin{align}\label{eq:outcome_eq}
  Y = Y_1 \cdot D + Y_0 \cdot (1-D)~.
\end{align}
In addition, following the framework of \cite{imbens/angrist:94}, let $Z$ denote the observed binary instrument, $D_0$ denote the potential value of treatment receipt if $Z=0$ and $D_1$ denote the potential value of treatment receipt if $Z=1$, i.e. the observed and potential treatment receipts are related by the equation
\begin{align}\label{eq:treatment_eq}
  D = D_1 \cdot Z + D_0 \cdot (1-Z)~.
\end{align}
Here the instrument is exogenous and, hence, jointly statistically independent of all the potential variables, i.e.
\begin{align}\label{eq:exclusion}
  (Y_0,Y_1, D_0, D_1) \perp Z~.
\end{align}
Consider the instrument monotonicity assumption of \cite{imbens/angrist:94}, i.e.
\begin{align}\label{eq:monotonicity}
  D_1 \geq D_0~,
\end{align}
which states that the value of the instrument monotonically affects the receipt of treatment for every unit in the population. Then, under the setup introduced above, I characterize and compare the identified sets, i.e. the set of all possible parameter values that are compatible with the distribution of the data and the imposed assumptions, for the distribution of potential outcomes, i.e.
\begin{align*}
  \text{Prob}\{(Y_0,Y_1) \in \mathcal{A}\}
\end{align*}
for all $\mathcal{A} \subseteq \{0,1\}^2$, in the absence and presence of the instrument monotonicity assumption. In particular, I study if and under what conditions can the latter identified set be a strict subset of the former given which we can then conclude that instrument monotonicity has identifying content on the distribution of potential outcomes.

Since instrument monotonicity additionally restricts the model, it is potentially possible that it has identifying content on the distribution of potential outcomes. As further elaborated in Remark \ref{rem:balke/pearl} below, \cite{balke/pearl:97}, \cite{heckman/vytlacil:01} and \cite{kitagawa:09} have previously explored this possibility on various features of the  distribution of potential outcomes. In the context of the setup described above, their conclusions can be generally summarized as follows. First, they find that imposing instrument monotonicity can misspecify the model, i.e. the distribution of the data may not be compatible with the imposed assumptions. Second, when the model is not misspecified, they find that instrument monotonicity has no identifying content on the marginal distributions of the potential outcomes, i.e.
\begin{align*}
  \left( \text{Prob}\{Y_0 \in \mathcal{A}_0\}, \text{Prob}\{Y_1 \in \mathcal{A}_1\} \right)
\end{align*} 
for all $\mathcal{A}_0,\mathcal{A}_1 \subseteq \{0,1\}$, in the sense that the identified sets for the marginal distributions of the potential outcomes with and without instrument monotonicity are identical. However, as there may exist distributions of potential outcomes that do not fall in the identified set for the distribution of potential outcomes but have marginals that fall in the identified set for the marginal distributions of the potential outcomes, the second conclusion does not allow us to also conclude that instrument monotonicity has no identifying content on the distribution of potential outcomes.

In this paper, I demonstrate that instrument monotonicity can indeed have identifying content on the distribution of potential outcomes when the model is not misspecified. Specifically, by characterizing and comparing the identified sets for the distribution of potential outcomes with and without instrument monotonicity, I find that they are not identical and that, for some distributions of the data, the identified set with instrument monotonicity can be a strict subset of that without instrument monotonicity. I derive necessary and sufficient conditions on the distribution of the data for when this is the case. As further discussed below, these conditions reveal two intuitive facts for instrument monotonicity to have identifying content. First, they reveal that there should be some so-called two-sided noncompliance observed in the data, i.e. 
\begin{align*}
  \text{Prob}\{D = 1 | Z = 0 \}  > 0~\text{ and }~\text{Prob}\{D = 0 | Z = 1 \}  > 0~,
\end{align*}
so that instrument monotonicity is an actual restriction and not automatically imposed by the data. Second, given there is such noncompliance observed in the data, they reveal that the additional restrictions that instrument monotonicity then implies on the distribution of potential outcomes should not be directly implied by those that logically arise from the definition of a probability distribution.

\section{Identification Analysis}

I begin by introducing the formal setup and notation for the analysis pursued in this paper. Let
\begin{align*}
  X = (Y,D,Z)
\end{align*}
denote the random variable that summarizes the observed random variables, which takes values on a discrete sample space $\mathcal{X} = \{0,1\}^3$. Similarly, let
\begin{align*}
 W =  (Y_0,Y_1,D_0,D_1)
\end{align*}
denote the random variable that summarizes the underlying random variables in the model, which takes values on a discrete sample space $\mathcal{W} = \{0,1\}^4$. Due to the discrete sample space, the distribution of the underlying random variables can be characterized by a probability mass function $Q$ with support contained in $\mathcal{W}$, i.e. $Q:\mathcal{W} \to [0,1]$ such that
\begin{align*}
  \sum_{w \in \mathcal{W}} Q(w) = 1~.
\end{align*}
Assume that the underlying random variables are related to the observed random variables by the equations in \eqref{eq:outcome_eq} and \eqref{eq:treatment_eq}. In addition, assume that the observed variable $Z$ is nondegenerate and statistically independent of the underlying random variables as stated in \eqref{eq:exclusion}. These assumptions together imply that $Q$ is related to the distribution of the data by
\begin{align}\label{eq:data_restriction}
  \sum_{w \in \mathcal{W}_x} Q(w) = \text{Prob}\{Y=y,D=d|Z=z\} \equiv P_{y,d|z}
\end{align}
for each $x = (y,d,z) \in \mathcal{X}$, where $\mathcal{W}_x$ is the set of all $w = (y_0,y_1,d_0,d_1) \in \mathcal{W}$ such that $d = d_1 \cdot z + d_0 \cdot (1-z)$ and $y = y_1 \cdot d + y_0 \cdot (1-d)$~. 

The analysis aims to study whether we can additionally learn something for a pre-specified parameter of interest based on $Q$ over the restrictions imposed on $Q$ in \eqref{eq:data_restriction} when the instrument monotonicity assumption stated in \eqref{eq:monotonicity} is also imposed. In particular, the instrument monotonicity assumption can be equivalently restated as a restriction on $Q$ as
\begin{align}\label{eq:mon_restriction}
  \text{Prob}\{D_0 = 1, D_1 = 0\} \equiv \sum\limits_{w \in \mathcal{W}_M} Q(w) = 0~,
\end{align}
where $\mathcal{W}_M = \{(y_0,y_1,d_0,d_1,z) \in \mathcal{W} : d_0=1, d_1=0\}$~. The parameter of interest I study corresponds to the distribution of potential outcomes. Formally, this parameter can be written as the following function of $Q$:
\begin{align}\label{eq:parameter}
  \bar{\mu}(Q)(\mathcal{A}) = \sum_{w \in \mathcal{W}_\mathcal{A}} Q(w) \equiv \text{Prob}\{(Y_0,Y_1) \in \mathcal{A}\}
\end{align}
for each $\mathcal{A} \in \mathcal{P}(\{0,1\}^2)$~, where $\mathcal{P}(\{0,1\}^2) = \{\mathcal{S} : \mathcal{S} \subseteq \{0,1\}^2\}$ and $\mathcal{W}_{\mathcal{A}} = \{(y_0,y_1,d_0,d_1,z) \in \mathcal{W} : (y_0,y_1) \in \mathcal{A}\}$. 

Using the above introduced setup, the objective of the analysis can formally be stated in terms of characterizing and comparing the identified sets, i.e. the set of feasible parameter values such that $Q$ satisfies the imposed restrictions, in the presence and absence of the restriction imposed by instrument monotonicity. To be specific, let $\mathbf{M}$ denote the set of all functions from $\mathcal{P}(\{0,1\}^2)$ to $\mathbf{R}$ and let $\mathbf{Q}$ denote the set of all probability mass functions with support contained in $\mathcal{W}$. Then, for the parameter defined in \eqref{eq:parameter}, I study the following two identified sets
\begin{align}
  \Theta_{E} &= \left\{ \mu \in \mathbf{M} : \mu = \bar{\mu}(Q) \text{ for some } Q \in \mathbf{Q}_{E} \right\}~, \label{eq:theta_E}\\
  \Theta_{EM} &= \left\{ \mu \in \mathbf{M} : \mu = \bar{\mu}(Q) \text{ for some } Q \in \mathbf{Q}_{EM} \right\}~, \label{eq:theta_EM}
\end{align}
where 
\begin{align*}
  \mathbf{Q}_{E} = \{ Q \in \mathbf{Q} : Q \text{ satisfies \eqref{eq:data_restriction}} \}
\end{align*}
denotes the set of all $Q \in \mathbf{Q}$ that satisfy the restrictions imposed by the data and instrument exogeneity, and
\begin{align*}
  \mathbf{Q}_{EM} = \{ Q \in \mathbf{Q} : Q \text{ satisfies \eqref{eq:data_restriction}  and  \eqref{eq:mon_restriction}} \}
\end{align*}
denotes the set of all $Q \in \mathbf{Q}$ that satisfy the restrictions imposed by the data, instrument exogeneity and instrument monotonicity. Note that since $\mathbf{Q}_{EM} \subseteq \mathbf{Q}_{E}$ as $\mathbf{Q}_{EM}$ is additionally restricted by instrument monotonicity, we always have that
\begin{align*}
  \Theta_{EM} \subseteq  \Theta_{E}~.
\end{align*}
However, since $\mathbf{Q}_{EM} \subsetneq \mathbf{Q}_{E}$ whenever $\mathbf{Q}_{E}$ is nonempty, it is then potentially possible that we can have
\begin{align*}
  \Theta_{EM} \subsetneq  \Theta_{E}~,
\end{align*}
in which case, we can conclude that instrument monotonicity has identifying content on the distribution of potential outcomes. The objective of the analysis presented below is to characterize $\Theta_{E}$ and $\Theta_{EM}$ and study if and when this can be the case.

Before proceeding to this analysis, note that it is potentially possible that $\mathbf{Q}_{E}$ or $\mathbf{Q}_{EM}$ can be empty, i.e. the model can be misspecified, when the distribution of the data is incompatible with the imposed assumptions on the model. As a result, if $\mathbf{Q}_{EM}$ is empty but $\mathbf{Q}_{E}$ is not, imposing instrument monotonicity can have identifying content as in this case it follows that
\begin{align*}
  \Theta_{EM} = \emptyset \subsetneq  \Theta_{E}~.
\end{align*}
As previously noted in \cite{balke/pearl:97} and \cite{kitagawa:09}, this case is indeed possible for some distributions of the data. In this paper, I focus on the identifying content of instrument monotonicity in the case where $\mathbf{Q}_{EM}$ and, hence, $\mathbf{Q}_{E}$ are nonempty. To this end, I suppose that $P$ satisfies
\begin{align}\label{eq:Pass} 
  \min\left\{P_{y,1|1} - P_{y,1|0}, P_{y,0|0} - P_{y,0|1}\right\} \geq 0~\text{ for each $y \in \{0,1\}$}~. 
  \end{align}
From \citet[][Proposition 1]{kitagawa:15}, it follows that these conditions on $P$ are necessary and sufficient to ensure that $\mathbf{Q}_{EM}$ is nonempty. 

The following proposition now states the result that respectively characterizes the identified sets in \eqref{eq:theta_E} and \eqref{eq:theta_EM}. The proof of this proposition is presented in Appendix \ref{sec:proof_JD}.

\begin{proposition}\label{th:JD}
Suppose that $P$ satisfies \eqref{eq:Pass}. Then $\Theta_E$ in \eqref{eq:theta_E} is equal to the set of all $\mu \in \mathbf{M}$ such that
\begin{align}
  \mu(\mathcal{A}) &= \sum_{a \in \mathcal{A}} \mu(\{a\})~, \label{eq:Erest1}\\  
  \mu(\mathcal{A}) &\in [L_{E}(\mathcal{A}), U_{E}(\mathcal{A})]~\label{eq:Erest2}
\end{align}  
for all $\mathcal{A} \subseteq \{0,1\}^2$, and $\Theta_{EM}$ in \eqref{eq:theta_EM} is equal to the set of all $\mu \in \mathbf{M}$ such that
\begin{align}
  \mu(\mathcal{A}) &= \sum_{a \in \mathcal{A}} \mu(\{a\})~, \label{eq:EMrest1}\\  
  \mu(\mathcal{A}) &\in [L_{EM}(\mathcal{A}) , U_{EM}(\mathcal{A})]~\label{eq:EMrest2}
\end{align}  
for all $\mathcal{A} \subseteq \{0,1\}^2$, where
\begin{align*}
\begin{array}{ l c l c l  } 
U_{E}(\emptyset) & = & U_{EM}(\emptyset) &=& 0~, \\
U_{E}(\{(i,j)\}) & = & U_{EM}(\{(i,j)\}) &=& \min\left\{P_{i,0|0} + P_{j,1|0},P_{i,0 |1} + P_{j,1|1}\right\}~, \\
U_{E}(\{(i,0),(i,1)\}) & = & U_{EM}(\{(i,0),(i,1)\}) &=& P_{i,0|0} + P_{0,1|0} + P_{1,1|0} ~,\\
U_{E}(\{(0,i),(1,i)\}) & = & U_{EM}(\{(0,i),(1,i)\}) &=& P_{i,1|1} + P_{0,0|1} + P_{1,0|1} ~,\\
U_{E}(\{(0,i),(1,1-i)\}) & = & U_{EM}(\{(0,i),(1,1-i)\}) &=& \min \{1, P_{0,0|0} + P_{1,0|1} + P_{i,1|0} + P_{1-i,1|1}, \\
& & & & P_{0,0|1} + P_{1,0|0} + P_{i,1|1} + P_{1-i,1|0} \}~, \\
\multicolumn{3}{c}{U_{E}(\{(i,j),(i,1-j),(1-i,j)\})} &=& \min \{  1, P_{i,0|0} + P_{1-i,0|1} + P_{1-j,1|0} + P_{j,1|1}  \\
 & & & & + \min\left\{ P_{i,0|1}, P_{j,1|0} \right\}  \} ~,\\
\multicolumn{3}{c}{U_{EM}(\{(i,j),(i,1-j),(1-i,j)\})} &=& \min \{  1, P_{i,0|0} + P_{1-i,0|1} + P_{1-j,1|0} + P_{j,1|1}   \} ~,\\
U_{E}\left(\{0,1\}^2\right) & = & U_{EM}\left(\{0,1\}^2\right) &=& 1~
\end{array}
\end{align*}
for all $i,j \in \{0,1\}$, and 
\begin{align*}
  L_{E}(\mathcal{A}) = 1 - U_{E}\left(\{0,1\}^2 \setminus\mathcal{A}\right)~\text{ and }~L_{EM}(\mathcal{A}) = 1 - U_{EM}\left(\{0,1\}^2 \setminus\mathcal{A}\right)~
\end{align*}
for all $\mathcal{A} \subseteq \{0,1\}^2$. Furthermore, for each given $\mathcal{A} \subseteq \{0,1\}^2$, $\Theta_{E}$ and $\Theta_{EM}$ are such that for every $t \in [L_{E}(\mathcal{A}) , U_{E}(\mathcal{A})]$ and $t' \in [L_{EM}(\mathcal{A}) , U_{EM}(\mathcal{A})]$, there exist $\mu \in \Theta_E$ and $\mu' \in \Theta_{EM}$ with $\mu(\mathcal{A}) = t$ and $\mu'(\mathcal{A}) = t'$.
\end{proposition}
  
Proposition \ref{th:JD} reveals that the identified sets with and without instrument monotonicity correspond to the set of all probability distributions defined on the sample space $\{0,1\}^2$ such that the probability assigned to any $\mathcal{A} \subseteq \{0,1\}^2$ is restricted to take values in some known closed interval. In addition, it reveals that these intervals are sharp for any $\mathcal{A} \subseteq \{0,1\}^2$ in the sense that for any value in these intervals there exists a distribution in the identified set with probability assigned to the event $\mathcal{A}$ equal to that value.

Given the above characterizations of $\Theta_{E}$ and $\Theta_{EM}$, we can next compare them to study if and when it is possible that the latter identified set is a strict subset of the former. To this end, observe that if 
\begin{align}\label{eq:bounds_strict}
  U_{EM}(\{(i,j),(i,1-j),(1-i,j)\}) < U_{E}(\{(i,j),(i,1-j),(1-i,j)\})~
\end{align}
holds for some $i,j \in \{0,1\}$, then it must be the case that $\Theta_{EM} \subsetneq \Theta_{E}$. In particular, this follows directly from Proposition \ref{th:JD} as since the intervals for the probability of any event are sharp, there exists a $\mu \in \Theta_{E}$ with
\begin{align*}
  \mu(\{(i,j),(i,1-j),(1-i,j)\}) \leq U_{E}(\{(i,j),(i,1-j),(1-i,j)\})~,
\end{align*}
and
\begin{align*}
 U_{EM}(\{(i,j),(i,1-j),(1-i,j)\}) < \mu(\{(i,j),(i,1-j),(1-i,j)\})
\end{align*}
for some $i,j \in \{0,1\}$, and hence such that $\mu \notin \Theta_{EM}$. In addition, it holds that the condition in \eqref{eq:bounds_strict} is in fact also necessary to ensure that $\Theta_{EM} \subsetneq \Theta_{E}$. To see why, note that if $\Theta_{EM} \subsetneq \Theta_{E}$, i.e. there exists a $\mu \in \Theta_{E}$ such that $\mu \notin \Theta_{EM}$, it must follow from the characterization of $\Theta_{E}$ and $\Theta_{EM}$ that for some $i,j \in \{0,1\}$ we have 
\begin{align*}
  \mu(\{(i,j),(i,1-j),(1-i,j)\}) \leq U_{E}(\{(i,j),(i,1-j),(1-i,j)\})~
\end{align*}
with
\begin{align*}
  U_{EM}(\{(i,j),(i,1-j),(1-i,j)\}) < \mu(\{(i,j),(i,1-j),(1-i,j)\})~,
\end{align*}
from which the condition in \eqref{eq:bounds_strict} then follows. Using the expressions for the bounds of the intervals in Proposition \ref{th:JD}, we can then rewrite the condition in \eqref{eq:bounds_strict} in a straightforward manner in terms of the distribution of the data. The following corollary presents the resulting conditions.

\begin{corollary}\label{cor:content}
Suppose that $P$ satisfies \eqref{eq:Pass}. Then, for $\Theta_{E}$ and $\Theta_{EM}$ respectively defined in \eqref{eq:theta_E} and \eqref{eq:theta_EM}, we have that
\begin{align*}
  \Theta_{EM} \subsetneq \Theta_{E}
\end{align*}
if and only if 
\begin{align}\label{eq:cond1}
  \min\left\{ P_{i,0|1}, P_{j,1|0} \right\} > 0
\end{align}
and
\begin{align}\label{eq:cond2}
  P_{i,0|0} + P_{1-i,0|1} + P_{1-j,1|0} + P_{j,1|1} < 1 
\end{align}
hold for some $i,j \in \{0,1\}$.
\end{corollary}

In order to better understand the above corollary, it useful to expand on what the two conditions in \eqref{eq:cond1} and \eqref{eq:cond2} intuitively capture. To this end, given that $P_{y,d|z} \geq 0$ logically holds for all $y,d,z \in \{0,1\}$, note that \eqref{eq:cond1} is satisfied for some $i,j \in \{0,1\}$ if and only if 
\begin{align*}
 \min \left\{ P_{0,0|1} + P_{1,0|1}, P_{0,1|0} + P_{1,1|0} \right\} > 0
\end{align*}
or, equivalently,
\begin{align*}
\min \left\{ \text{Prob} \{ D = 0 | Z = 1 \}, \text{Prob} \{ D = 1 | Z = 0 \} \right\} > 0
\end{align*}
holds, i.e. the probability of units whose received treatment does not equal the instrument in each group determined by the instrument value is strictly greater than zero, often referred to as two-sided noncompliance. In turn, since \eqref{eq:data_restriction} implies
\begin{align*}
   \text{Prob} \{ D = 0 | Z = 1 \} &= \text{Prob}\{D_0 = 1, D_1 = 0\} + \text{Prob}\{D_0 = 0, D_1 = 0\}~, \\ 
   \text{Prob} \{ D = 1 | Z = 0 \} &= \text{Prob}\{D_0 = 1, D_1 = 0\} + \text{Prob}\{D_1 = 1, D_1 = 1\}~, 
\end{align*}
it follows that absent this condition, the distribution of the data itself implies the instrument monotonicity restriction in \eqref{eq:mon_restriction} and, hence, additionally imposing it does not restrict $Q$ in any additional manner. As a result, the first condition captures the fact that for instrument monotonicty to have identifying content we need that it is not automatically implied by the distribution of the data. 

Given this condition holds and hence that instrument monotonicity actually imposes restrictions on $Q$, note from Proposition \ref{th:JD} that the additional restrictions weakly reduce the upper bound on the probability that the potential outcomes lie in $\{(i,j),(i,1-j),(1-i,j)\}$, or equivalently the lower bound on the probability that they lie in $\{(1-i,1-j)\}$, for some $i,j \in \{0,1\}$. However, for these restrictions to strictly reduce the upper bound and hence actually result in identifying content, we need that
\begin{align*}
  U_{EM}(\{(i,j),(i,1-j),(1-i,j)\}) < 1
\end{align*}
as otherwise the upper bound is implied by the fact that the probability of any event must be bounded above by one. This gives rise to the condition in \eqref{eq:cond2}. In other words, the second condition captures the fact that for the additional restrictions introduced by instrument monotoncitiy to have identifying content, we need that the additional information they imply on the distribution of potential outcomes is not directly implied by the logical restrictions that arise from $Q$ being a probability mass function.

\begin{remark}
Following results in \cite{vytlacil:02}, note that the instrument monotonicity assumption in \eqref{eq:monotonicity} is equivalent to assuming a separable treatment selection equation, i.e.
\begin{align*}
  D_z = 1\{ z \cdot \beta + \epsilon \geq 0 \}~,
\end{align*}
where $\beta$ is a nonrandom and nonnegative real number. In turn, the analysis in this paper equivalently studies the identifying content of imposing a separable treatment selection equation.
\end{remark}

\begin{remark}\label{rem:proof_strategy}
The proof of Proposition \ref{th:JD} begins by exploiting a linear programming strategy from \cite{balke/pearl:97} to obtain bounds on the probability that the distribution of potential outcomes assigns to each event $\mathcal{A} \subseteq \{0,1\}^2$, presented in \eqref{eq:Erest2} and \eqref{eq:EMrest2}. It then proceeds to show that these bounds are jointly sharp for the distribution of potential outcomes by showing that for any $\mu \in \mathbf{M}$ satisfying \eqref{eq:Erest1} and \eqref{eq:Erest2} (or \eqref{eq:EMrest1} and \eqref{eq:EMrest2}) for all $\mathcal{A} \subseteq \{0,1\}^2$, there exists a $Q$ in $\mathbf{Q}_{E}$ (or $\mathbf{Q}_{EM}$) such that $\mu = \bar{\mu}(Q)$. In order to show this, the proof exploits an alternative linear programming strategy from \cite{bai/etal:19}, who use this strategy in related settings to derive, amongst other results, conditions on the distribution of the data for when the model is misspecified---see also \cite{machado/etal:18} for a related strategy.
\end{remark}

\begin{remark}
 In a model without potential treatment receipts, \cite{beresteanu/etal:12} characterize the identified set for the distribution of potential outcomes under the assumption that the instrument is statistically independent of the potential outcomes, i.e.
 \begin{align*}
   (Y_0,Y_1) \perp Z~.
 \end{align*}
When $P$ satisfies \eqref{eq:Pass}, their resulting identified set is identical to $\Theta_E$ stated in Proposition \ref{th:JD}. See also \cite{richardson/robins:10} for related results on the distribution of potential outcomes conditional on the event $\{D_0 = d_0, D_1 = d_1\}$ for each $d_0,d_1 \in \{0,1\}$.
\end{remark}

\begin{remark}\label{rem:balke/pearl}
The marginal distributions of the potential outcomes are determined by the following function of $Q$:
\begin{align*}
  \mu^{\dagger}(Q)(\mathcal{A}_0,\mathcal{A}_1) = \left( \sum_{w \in \mathcal{W}_{\mathcal{A}_0}}Q(w), \sum_{w \in \mathcal{W}_{\mathcal{A}_0}}Q(w)   \right) \equiv \left(\text{Prob}\{Y_0 \in \mathcal{A}_0\},\text{Prob}\{Y_1 \in \mathcal{A}_1\}\right)
\end{align*} 
for each $\mathcal{A}_0,\mathcal{A}_1 \subseteq \mathcal{Y}$, where $\mathcal{W}_{\mathcal{A}_d} = \{(y_0,y_1,d_0,d_1) \in \mathcal{W} : y_d \in \mathcal{A}_d  \}$ for $d \in \{0,1\}$. When the model is not misspecified, \cite{kitagawa:09} shows that instrument monotonicity has no identifying content for this function. In turn, it also directly follows that in this case instrument monotonicity has no identifying content for any function based on the marginal distributions of the potential outcomes such as the average treatment effect:
  \begin{align*}
    E[Y_1 - Y_0] = \text{Prob}\{Y_1 = 1\} - \text{Prob}\{Y_0 = 1\}~.
  \end{align*}
\cite{balke/pearl:97} and \cite{heckman/vytlacil:01} also provide results focusing on specific  functions of the marginal distributions of the potential outcomes such as the average treatment effect. Note that \cite{heckman/vytlacil:01} and \cite{kitagawa:09} also respectively extend their analyses to cases where the outcomes and instruments are continuous and where the outcomes are continuous.
\end{remark}

\begin{remark}
A direct consequence of Corollary \ref{cor:content} and previous results noted in Remark \ref{rem:balke/pearl} is the conclusion that, when the model is not misspecified, imposing instrument monotonicity does not have any identifying content for parameters based on the marginal distributions of the potential outcomes, but can have identifying content for those based only on the joint distribution. Example of the latter class of parameters include those such as 
\begin{align*}
 \text{Prob}\{Y_1 > Y_0\} &= \bar{\mu}(Q)(\{(0,1)\})~, \\
  \text{Prob}\{Y_0 > Y_1\} &= \bar{\mu}(Q)(\{(1,0)\})~,
\end{align*}
which respectively correspond to the proportion who strictly benefit and lose---see \cite{heckman/etal:97} and \cite{manski:97} for an early exposition on the policy relevance of such parameters based on the joint distribution of the potential outcomes. Note that since each of the two examples correspond to the probability that the potential outcomes lie in some set $\mathcal{A} \subseteq \{0,1\}^2$, Proposition \ref{th:JD} allows us to directly characterize their identified sets with and without instrument monotonicity, respectively given by the intervals $[L_{EM}(\mathcal{A}) , U_{EM}(\mathcal{A})]$ and $[L_{E}(\mathcal{A}) , U_{E}(\mathcal{A})]$~.
\end{remark}

\newpage
\appendix
\renewcommand{\theequation}{\Alph{section}-\arabic{equation}}

\section{Proof of Proposition \ref{th:JD}}\label{sec:proof_JD}

To prove the first assertion of the proposition, we need to show that $\mu \in \Theta_E$ if and only if $\mu \in \mathbf{M}$ and $\mu$ satisfies \eqref{eq:Erest1} and \eqref{eq:Erest2} for each $\mathcal{A} \subseteq \{0,1\}^2$; and that $\mu \in \Theta_{EM}$ if and only if $\mu \in \mathbf{M}$ and $\mu$ satisfies \eqref{eq:EMrest1} and \eqref{eq:EMrest2} for each $\mathcal{A} \subseteq \{0,1\}^2$. Below, I present the proofs of the if, and only if parts of the statements in two separate parts.

\textbf{Part 1}: If $\mu \in \Theta_{E}$, it implies that there exists a $Q \in \mathbf{Q}_E$ such that $\mu = \bar{\mu}(Q)$. Note that $\bar{\mu}(Q) \in \mathbf{M}$. It remains to show that $\bar{\mu}(Q)$ satisfies \eqref{eq:Erest1} and \eqref{eq:Erest2} for each $\mathcal{A} \subseteq \{0,1\}^2$. To this end, recall that
\begin{align}\label{eq:parameter_structure}
  \bar{\mu}(Q)(\mathcal{A}) = \sum_{w \in \mathcal{W}_{\mathcal{A}}} Q(w)~,
\end{align}
for each $\mathcal{A} \subseteq \{0,1\}^2$, where $\mathcal{W}_{\mathcal{A}}$ is defined in \eqref{eq:parameter}, and that $Q$ is required to satisfy the following restrictions
\begin{align}
  0 \leq Q(w) &\leq 1~ \text{ for each }~w \in \mathcal{W}~ , \label{eq:rest1} \\
  \sum_{w \in \mathcal{W}} Q(w) &= 1~, \label{eq:rest2}
\end{align}
along with those in \eqref{eq:data_restriction}. From the definition of the function in \eqref{eq:parameter_structure}, it directly follows that $\bar{\mu}(Q)$ satisfies \eqref{eq:Erest1} as
\begin{align*}
 \bar{\mu}(Q)(\mathcal{A}) = \sum_{a \in \mathcal{A}} \sum_{w \in \mathcal{W}_a} Q(w) = \sum_{a \in \mathcal{A}} \bar{\mu}(Q)(\{a\})
\end{align*}
for each $\mathcal{A} \subseteq \{0,1\}^2$. Next, to show that it satisfies \eqref{eq:Erest2}, note that $\mathbf{Q}_E$ is a closed and convex set, given the linear restrictions it is determined by in \eqref{eq:rest1}, \eqref{eq:rest2} and \eqref{eq:data_restriction}, and also nonempty, given $P$ satisfies \eqref{eq:Pass}, and that the function in \eqref{eq:parameter_structure} is a real-valued continuous function of $Q$ for each $\mathcal{A} \subseteq \{0,1\}^2$. As a result, it follows that
\begin{align*}
  \bar{\mu}(Q)(\mathcal{A}) \in [ \mu_{L}(\mathcal{A}), \mu_U(\mathcal{A}) ]~,
\end{align*}
for each $\mathcal{A} \subseteq \{0,1\}^2$, where
\begin{align}\label{eq:opt_problems}
\mu_L(\mathcal{A}) = \min_{\{Q(w)\}_{w \in \mathcal{W}}} \sum_{w \in \mathcal{W}_{\mathcal{A}}} Q(w)~ \text{ and }~\mu_U(\mathcal{A}) = \max_{\{Q(w)\}_{w \in \mathcal{W}}} \sum_{w \in \mathcal{W}_{\mathcal{A}}} Q(w)~
\end{align}
subject to $Q$ satisfying the restrictions in \eqref{eq:rest1}, \eqref{eq:rest2} and \eqref{eq:data_restriction}. Using a procedure introduced in \cite{balke/pearl:97}, we can obtain symbolic expressions for the optimal values of these two optimization problems for each $\mathcal{A} \subseteq \{0,1\}^2$. For completeness, I next describe this procedure for a given $\mathcal{A} \subseteq \{0,1\}^2$. For notational simplicity, let $q$ denote a vector capturing $(Q(w) : w \in \mathcal{W})$ in vector notation, and let $A$ denote a matrix and $c$ and $b$ denote vectors such that the problems in \eqref{eq:opt_problems} can equivalently be rewritten in matrix notation as
\begin{align*}
  \mu_L(\mathcal{A}) = \min_{q} c'q~\text{ and }~\mu_U(\mathcal{A}) = \max_q c'q~
\end{align*}
subject to the constraint that $Aq \leq b ,~q \geq 0$~. By the strong duality theorem, note that the optimal values of these two optimization problems are equal to that of their dual problems, i.e.
\begin{align*}
  \mu_L(\mathcal{A}) = \max_{u} b'u~ \text{ and }~\mu_U(\mathcal{A}) = \min_{u} b'u~
\end{align*}
subject to the constraint that $A'u \geq c,~ u \geq 0$~. Since the optimal solution of a linear programming problem exists at a vertex of the constraint set, it then follows that
\begin{align*}
  \mu_L(\mathcal{A}) = \max_{v \in \mathcal{V}} b'v~\text{ and }~\mu_U(\mathcal{A}) = \min_{v \in \mathcal{V}} b'v~,
\end{align*}
where $\mathcal{V}$ denotes the set of vertices of the constraint set $\left\{u \in \mathbf{R}^K : A'u \geq c,~ u \geq 0 \right\}$ such that $K$ denotes the row dimension of the matrix $A$~. By exploiting the fact that the constraint set is known, we can use a vertex enumeration algorithm to computationally characterize $\mathcal{V}$ and, in turn, obtain expressions for $\mu_L(\mathcal{A})$ and $\mu_U(\mathcal{A})$. Applying this procedure and then simplifying these expressions by taking into account that $P$ satisfies \eqref{eq:Pass} as well as
\begin{align}
  P_{y,d|z}  \in [0,1]~&~\text{ for }~y,d,z \in \{0,1\}~, \label{eq:Pcond1} \\
  \sum_{y,d \in \{0,1\}} P_{y,d|z}  = 1 ~&~\text{ for }~ z \in \{0,1\}~, \label{eq:Pcond2}
\end{align}
given by the fact that it is a conditional probability distribution, one obtains that $\mu_L(\mathcal{A}) = L_{E}(\mathcal{A})$ and $\mu_U(\mathcal{A}) = U_{E}(\mathcal{A})$ for each $\mathcal{A} \subseteq \{0,1\}^2$ and, in turn, that $\bar{\mu}(Q)$ satisfies \eqref{eq:Erest2}. An analogous argument can be used to show that if $\mu \in \Theta_{EM}$ then $\mu \in \mathbf{M}$ and $\mu$ satisfies \eqref{eq:EMrest1} and \eqref{eq:EMrest2}, where in the procedure described above to obtain symbolic expressions for the bounds, the additional restriction in \eqref{eq:mon_restriction} is also introduced. This completes the first part of the proof for the first assertion of the proposition.

\textbf{Part 2}: Take $\mu \in \mathbf{M}$ satisfying \eqref{eq:Erest1} and \eqref{eq:Erest2} for each $\mathcal{A} \subseteq \{0,1\}^2$. To show that $\mu \in \Theta_E$, we need to show that there exists a $Q \in \mathbf{Q}_E$, i.e. a $Q$ satisfying the restrictions in \eqref{eq:rest1}, \eqref{eq:rest2} and \eqref{eq:data_restriction}, such that $\mu = \bar{\mu}(Q)$. We can show that such a $Q$ exists using a procedure introduced in \cite{bai/etal:19}. For completeness, I next describe this procedure. For notational simplicity, let $q$ denote, as before, a vector capturing $(Q(w) : w \in \mathcal{W})$ in vector notation, let $\tilde{A}_1$ and $\tilde{b}_1$ respectively denote a matrix and a vector such that the restrictions in \eqref{eq:rest1} and \eqref{eq:rest2} 
can equivalently be rewritten in matrix notation as
\begin{align}\label{eq:part2_matrix}
  \tilde{A}_1q \leq \tilde{b}_1~,
\end{align}
and let $\tilde{A}_2$ and $\tilde{b}_2$ respectively denote a matrix and a vector such the restrictions in \eqref{eq:data_restriction} and those imposed by $\mu = \bar{\mu}(Q)$ can equivalently be rewritten in matrix notation as
\begin{align*}
  \tilde{A}_2q = \tilde{b}_2~.
\end{align*}
Then, note that showing there exists a $Q$ satisfying the restrictions in \eqref{eq:rest1}, \eqref{eq:rest2} and \eqref{eq:data_restriction} such that $\mu = \bar{\mu}(Q)$ is equivalent to showing that 
\begin{align*}
  \tilde{b}_2 \in \left\{\tilde{A}_2 q : \tilde{A}_1 q \leq \tilde{b}_1,~ q \in \mathbf{R}^J \right\} \equiv \mathcal{B}~,
\end{align*}
where $J$ denotes the dimension of the vector $q$. \citet[][Lemma 2.1]{bai/etal:19} shows that
\begin{align*}
  \mathcal{B} = \text{conv}(\mathcal{S}_2)~,
\end{align*}
where 
\begin{align*}
  \mathcal{S}_2 = \left\{ \tilde{A}_2 v : v \text{ is an extreme point of the set } \mathcal{S}_1 \equiv \left\{ q \in \mathbf{R}^J : \tilde{A}_1 q \leq \tilde{b}_1 \right\} \right\}~
\end{align*}
and conv($\mathcal{S}_2$) denotes the convex hull of $\mathcal{S}_2$. In particular, this result allows us to use a computational algorithm to characterize $\mathcal{B}$ and, in turn, show $\tilde{b}_2 \in \mathcal{B}$. To see how, note that, similar to before, we can first use a vertex enumeration algorithm to collect the vertices of the set $\mathcal{S}_1$, which allows us to compute $\mathcal{S}_2$. Then, we can use a facet enumeration algorithm to obtain the so-called $\mathcal{H}$-representation of conv($\mathcal{S}_2$), which represents $\mathcal{B}$ in terms of a system of linear equality and inequality equations. Applying this procedure to obtain the $\mathcal{H}$-representation of $\mathcal{B}$, one finds that $\tilde{b}_2 \in \mathcal{B}$ if and only if $P$ satisfies \eqref{eq:Pcond1} and \eqref{eq:Pcond2} along with \eqref{eq:Pass} and $\mu$ satisfies \eqref{eq:Erest1} and \eqref{eq:Erest2} for each $\mathcal{A} \subseteq \{0,1\}^2$. Indeed, since $P$ and $\mu$ satisfy these restrictions, it follows that $\tilde{b}_2 \in \mathcal{B}$. An analogous argument can be used to show that if $\mu \in \mathbf{M}$ and satisfies \eqref{eq:EMrest1} and \eqref{eq:EMrest2} for each $\mathcal{A} \subseteq \{0,1\}^2$ then there exists a $Q \in \mathbf{Q}_{EM}$ such that $\mu = \bar{\mu}(Q)$, where in the procedure outlined above the additional restriction in \eqref{eq:mon_restriction} is also introduced in the restrictions captured by \eqref{eq:part2_matrix}. This completes the proof of the first assertion of the proposition.

In order to show the second assertion of the proposition, we can expand on certain elements of Part 1 above. Specifically, recall that since $\mathbf{Q}_E$ is a closed, convex and nonempty set and since $\bar{\mu}(Q)(\mathcal{A})$ is a continuous real-valued function of $Q$ for each $\mathcal{A} \subseteq \{0,1\}^2$, it follows that the image of this function over $\mathbf{Q}_E$, given by the set 
\begin{align}\label{eq:identifiedset_A}
  \left\{t \in \mathbf{R} : \bar{\mu}(Q)(\mathcal{A}) =t \text{ for some } Q \in \mathbf{Q}_E \right\}~,
\end{align}
is also a closed, convex and nonempty set. In particular, it follows that this set is a closed interval, where the lower and upper bounds are respectively given by
\begin{align*}
  \mu_L(\mathcal{A}) = L_E(\mathcal{A})~\text{ and }~\mu_U(\mathcal{A}) = U_E(\mathcal{A})~,
\end{align*}
where $\mu_L(\mathcal{A})$ and $\mu_U(\mathcal{A})$ are defined in \eqref{eq:opt_problems} and the second equality follows from Part 1 above. To complete the proof, note that it follows from the definition and characterization of the set in \eqref{eq:identifiedset_A} that for each $t \in [L_E(\mathcal{A}),U_E(\mathcal{A})]$, we have that there exists a $Q \in \mathbf{Q}_E$ and, in turn, a $\mu = \bar{\mu}(Q) \in \Theta_E$ such that $t = \bar{\mu}(Q)(\mathcal{A}) = \mu(\mathcal{A})$. An analogous argument can be used to show that for every $t' \in [L_{EM}(\mathcal{A}),U_{EM}(\mathcal{A})]$ there exists a $\mu' \in \Theta_{EM}$ such that $\mu'(\mathcal{A}) = t'$. This completes the proof of the second assertion of the proposition.

\newpage
\normalsize

\bibliography{Appendix/references.bib}


\end{document}